\documentclass[structabstract]{aa}  
\usepackage{graphicx}
\usepackage{color}
\usepackage{epsfig}
\usepackage{longtable,lscape}
\usepackage{subfigure}
\usepackage[varg]{txfonts}
\usepackage{natbib}
\bibpunct{(}{)}{;}{a}{}{,}  
\usepackage{rotating}

\def\kms {\rm{km~s^{-1}}}

\begin{document}
\title{Comparing galaxy populations in compact and loose groups of galaxies II: 
brightest group galaxies}
\author{H\'ector J. Mart\'inez, Valeria Coenda \& Hern\'an Muriel}
\institute{Instituto de Astronom\'ia Te\'orica y Experimental (IATE), 
CONICET-Observatorio Astron\'omico, Universidad Nacional de C\'ordoba. Laprida 854, 
C\'ordoba, X5000BGR. Argentina.\\
\email{julian@oac.uncor.edu}}
\date{Received  ; accepted  }
\abstract
{}
{The properties of the brightest galaxies (BCGs) are studied in both compact and loose 
groups of galaxies in order to better understand the physical mechanisms influencing 
galaxy evolution in different environments.}
{Samples of BCGs are selected in the compact groups identified by 
McConnachie et al. (2009), and in loose groups taken from Zandivarez \& Mart\'inez 
(2011). The following physical properties of the BCGs in compact groups and in subsamples
of loose groups are compared, defined by their mass and total luminosity: absolute 
magnitude, colour, size, surface brightness, stellar mass, concentration and 
morphological information from the Galaxy Zoo. The fraction of BCGs classified as red 
and/or early-type as a function of galaxy luminosity are studied. 
The fraction of the group's total luminosity contained in the
BCG and the difference in luminosity between the BCG and the second-ranked galaxy, are
also analysed.
}
{Some properties of BCGs in compact and loose groups are comparable. 
However, BCGs in compact groups are systematically more concentrated and have larger 
surface brightness than their counterparts in both, high- and low-mass loose groups.
The fractions of red and early-type BCGs in compact groups are consistent with those of
high-mass loose groups. Comparing BCGs in subsamples of compact and loose groups 
selected for their similar luminosities, BCGs in compact groups are found to be, on 
average, brighter, more massive, larger, redder and more frequently classified as 
elliptical. In compact groups, the BCG contains a larger fraction of the system's total 
luminosity and differs more in absolute magnitude from the second-ranked galaxy.
Using a simple model, which dry-merges the BCG in loose groups with a random choice
among the 2nd, 3rd and 4th-ranked galaxies in the group, and allowing for some star loss
in the process, we show that the absolute magnitude distributions of BCGs in compact 
and loose groups of similar luminosities can be made more alike.
}
{BCGs in compact and loose groups are found to be different. Some mechanisms 
responsible for transforming late-type galaxies into early types, such as mergers, may 
be more effective within compact groups due to their high densities and small velocity 
dispersion, which would lead their BCGs along somewhat different evolutionary paths.
}
\authorrunning{Mart\'inez, Coenda \& Muriel}
\titlerunning{Brightest group galaxies}
\maketitle
\section{Introduction} 
\label{sec:intro}
The brightest members of galaxy systems have been the subject of numerous studies. 
While most of them are concentrated in clusters of galaxies, the brightest galaxies in 
groups have been less studied, and even fewer studies have focused on the brightest 
members of compact groups. Brightest cluster galaxies (BCGs\footnote{The acronym will be 
used generically for both clusters and groups of galaxies}) are typically early-type and
are among the most massive galaxies known in the Universe. Typically, they have old 
stellar populations and represent a major challenge for models of galaxy formation and 
evolution. Cluster BCGs are known to have narrow luminosity and colour distributions 
(e.g. \citealt{postman:1995}). Situated in the densest and most extreme environments in 
the Universe, they may be not just examples of the bright end of the cluster luminosity 
function, but a different class of objects with their own luminosity function.  

It is generally assumed that BCGs acquire most of their stellar mass via dry mergers 
between smaller halos \citep{DeLucia07}, although the low evolution in stellar mass 
observed for BCGs at different redshifts (\citealt{Whiley08,Collins09,Stott10}) puts 
strong constraints on semi-analytic models in the $\Lambda$CDM cosmology (see for 
instance \citealt{Tonini12,Martizzi12}). The weak evolution of the stellar mass of BCGs 
with redshift suggests that a large fraction of mergers have happened at high redshift. 
Mergers at lower redshift are also expected, although their frequency may be a strong 
function of the environment. \citet{Liu09} studied the fraction of BCGs that show 
evidence of ongoing major dry mergers and found that this fraction increases with 
cluster richness. They suggest that BCG luminosity has increased on average by 15\% 
since $z\sim 0.7$ due to dry mergers. \citet{Edwards12} analyse a sample of BCGs that 
have close neighbours and conclude that mergers may have provided up to 10\% of the mass 
of BCGs since $z\sim0.3$. Similar results are found by \citet{Lidman12}, but they also 
suggest that a significant fraction of the mass involved in mergers is lost to the 
intra-cluster medium (see also \citealt{Conroy07} and \citealt{Stott10}). 

The basic question about BCGs is whether they are special or just the statistical 
result of selecting the brightest object of a given luminosity function (see 
\citealt{Paranjape12} and references therein). If mergers are important in the formation 
of BCGs, this should be reflected in the magnitude gap between the first- and 
second-ranked galaxies ($\Delta M_{12}\equiv M_2-M_1$). \citet{Smith10} computed 
$\Delta M_{12}$ in a sample of clusters and found a $3\sigma$ excess over the prediction
from Monte Carlo simulations of a Schechter function that fits the mean luminosity 
function of cluster galaxies. Although there is a general consensus that BCGs in 
clusters are special objects, there are some contradictory results for groups of 
galaxies. \citet{Geller:1983} studied a sample of groups of galaxies and concluded that 
their brightest members are less dominant than those in clusters and are consistent with
being the luminous tail of the luminosity function (see also \citealt{Lin:2010} and 
\citealt{Loh:2006}). Nevertheless, \citet{Paranjape12} found that BCG luminosity 
distributions in groups are inconsistent with these galaxies having been drawn from a 
universal luminosity function. 

Compact groups are a special type of groups of galaxies: even though they have velocity
dispersions comparable to those found in loose groups, they have higher densities, 
similar to those found in clusters (\citealt{Hickson92}), thus providing a different 
scenario for galaxy mergers. These exceptional conditions may significantly affect the 
evolution of BCGs in compact groups. The historically small size of the available samples
of compact groups has greatly limited statistical study of their brightest members.
\citet{Diaz12} used a sample of 78 compact groups and found a large magnitude gap between
the first- and second-ranked galaxy. However, they also found that this effect is not 
present in the samples of compact groups constructed by \citet{Hickson:1992}, 
\citet{Allam:2000} and \citet{Focardi:2002}.

Based on the sixth data release of the Sloan Digital Sky Survey (\citealt{dr6}), 
\citet{McConnachie:2009} identified a large sample of compact groups suitable for 
detailed statistical studies. With this sample, in \citet{PaperI} (hereafter Paper I) 
we compared the properties of galaxies in compact and loose groups, using large samples 
comprising 846 compact groups and 2536 (2528) loose groups of low (high) mass. We found 
significant differences between the mean properties of galaxies in groups, depending on 
whether they are in loose or in compact groups. We suggested that the physical mechanisms
that transform galaxies into earlier types may be more effective in compact than in loose
groups. 

This second article of the series is a comparative study of BCG properties in loose and 
compact groups in the context of the merger scenario. Given the controversial results in 
the literature for both loose and compact groups, its main purpose is to establish how 
differences in the dynamic properties of galaxy systems affect the formation and 
evolution of their brightest galaxies. We also explore whether the differences found in 
Paper I regarding the overall population of galaxies inhabiting loose and compact groups
are also present for BCGs.

The layout of the paper is as follows: in section \ref{sec:sample}, we describe the samples 
of groups and galaxies used; we perform a comparative analysis of the BCGs in compact and in 
loose groups in section \ref{sec:analysis}; finally, our results are summarised and discussed 
in section \ref{sec:discussion}. Throughout the paper, a flat cosmological model is assumed, 
with parameters $\Omega_0=0.3$, $\Omega_{\Lambda}=0.7$, and a Hubble's constant 
$H_0=100~h~\kms~{\rm Mpc}^{-1}$. All magnitudes were corrected for Galactic extinction 
using the maps by \citet{sch98} and are in the AB system. Absolute magnitudes and galaxy
colours were $K-$corrected using the method of \citet{Blanton:2003}~({\small KCORRECT} 
version 4.1).
\section{The samples of brightest group galaxies}
\label{sec:sample}
As in Paper I, the sample of compact groups used in this paper is drawn from catalogue A
of \citet{McConnachie:2009}, who used the original selection criteria of 
\citet{Hickson:1982} to identify compact groups in the sixth data release of the SDSS 
\citep{dr6}. Their catalogue A has 2,297 groups, adding up to 9,713 galaxies, down to a 
Petrosian (\citealt{petro76}) limiting magnitude of $r=18$, and has spectroscopic 
information for 4,131 galaxies (43\% completeness). In order to exclude interlopers, 
this catalogue includes only groups that have a maximum line-of-sight velocity difference
smaller than $1,000\, \kms$. For the purposes of this work, we selected all groups in the
A catalogue within the redshift range $0.05\le z\le0.18$ that have measured spectroscopic
redshift for their BCGs. This results in a sample of 477 compact groups.

The samples of loose groups used in this paper were drawn from the sample of groups 
identified by \citet{ZM11} in the Main Galaxy Sample \citep{Strauss:2002} of the seventh
data release of SDSS \citep{dr7} with $14.5 \le r \le 17.77$. They used a 
friends-of-friends algorithm \citep{H&G:1982} to link galaxies into groups, followed by a
second identification using a higher density contrast in groups with at least 10 members,
in order to split merged systems and clean up any spurious member detection. The authors
computed group virial masses from the virial radius of the systems and the velocity 
dispersion of member galaxies \citep{Limber:1960,Beers:1990}. The sample of \citet{ZM11}
comprises 15,961 groups with more than 4 members, adding up to 103,342 galaxies.
\begin{figure*}[t]
\centering
\includegraphics[width=19cm]{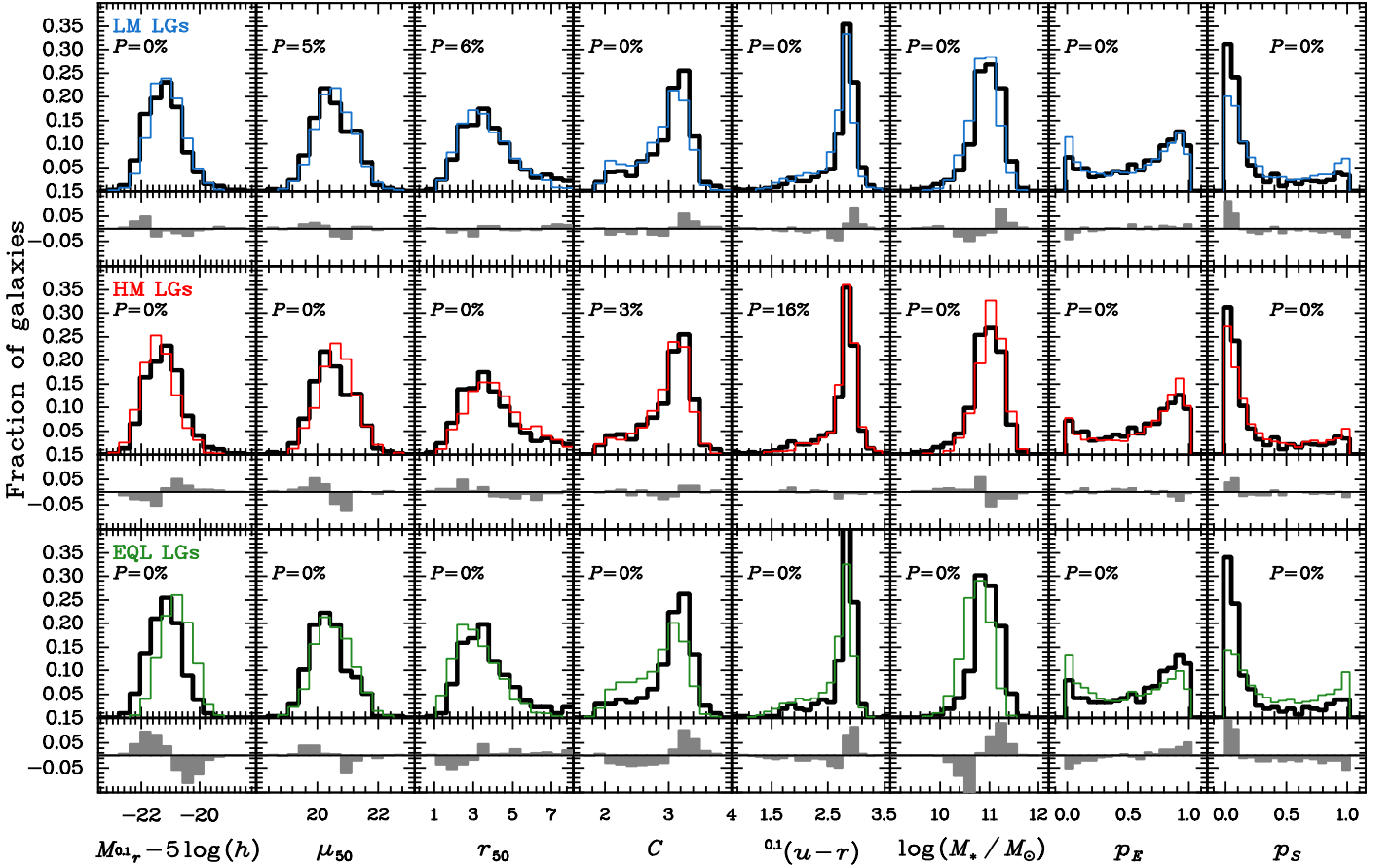}
\caption{Distributions of BCG properties, from {\em left} to {\em right}: 
$^{0.1}r-$band absolute magnitude; $r-$band surface brightness in 
${\rm mag\ arcsec^{-1}}$; Petrosian half-light radius in $h^{-1}{\rm kpc}$; concentration
parameter; $^{0.1}(u-r)$ colour; stellar mass; and the probability of being elliptical 
($p_E$) or spiral ($p_S$). In the {\em top} and {\em middle} rows, {\em thick black} line
corresponds to the sample of compact groups, {\em blue} line to low mass loose groups, 
and {\em red} line to high mass loose groups. The {\em bottom} row compares the equal 
luminosity subsamples of compact ({\em thick black}) and loose groups ({\em green}). All
distributions have been normalised to have the same area. Below each panel, we show as 
{\em shaded histograms} the residuals between the distributions. We quote in each panel 
the probability of the null hypothesis that both data sets are drawn from the same 
distribution according to the Kolmogorov-Smirnov test.
}
\label{fig:histo}
\end{figure*}
It is well-known that the properties of galaxies in groups are correlated with group mass
(e.g. \citealt{MM2:2006}). Thus, as in Paper I, we divided the groups in the \citet{ZM11}
sample into two subsamples of low, $\log(M/M_{\odot}h^{-1})\le 13.2$, and high, 
$\log(M/M_{\odot}h^{-1})\ge 13.6$, mass. To perform a fair comparison of BCGs, we used a
Monte Carlo algorithm to randomly select groups from these two subsamples, in order to 
construct new subsamples of low- and high-mass loose groups that have redshift 
distributions similar to that of the compact groups, in the redshift range 
$0.05\le z\le0.18$. The final subsamples of low- and high-mass loose groups include 1,414
and 1,458 systems, respectively.

Since compact groups in the \citet{McConnachie:2009} sample have no measured mass, and in
Paper I we were interested in comparing galaxies in samples of compact and loose groups 
that shared a similar physical magnitude, we used the group total luminosity as a common
parameter, and constructed samples of compact and loose groups that have similar 
luminosities. In this paper, we compare the properties of the BCGs of groups with similar
luminosity. Loose group luminosities were computed by \citet{Martinez:2012} and the 
luminosities of the compact groups were computed in Paper I, using the method of 
\citet{Moore:1993}. By means of the same Monte Carlo algorithm used in Paper I, we 
constructed two 'equal luminosity' subsamples of compact and loose groups of galaxies 
that have similar redshift and absolute magnitude distributions within the boundaries 
$0.05\le z\le 0.15$ and $-20.7\le M_{^{0.1}r}^{GROUP}-5\log(h)\le -23.8$. The equal 
luminosity subsamples of compact groups (EQL CGs) and loose groups (EQL LGs) include 314
and 1,577 systems, respectively. \\
\section{Comparing the brightest galaxies of compact and loose groups}
\label{sec:analysis}
\subsection{Galaxy properties}
\label{sec:comparing}
This study compares a number of galaxy parameters: 
\begin{itemize}
\item Petrosian absolute magnitude in the $^{0.1}r-$band;
\item the radius that encloses 50\% of the Petrosian flux, $r_{50}$;
\item the $r$-band surface brightness, $\mu_{50}$, computed inside $r_{50}$;
\item the $r-$band concentration index $C$, defined as the ratio of the radii enclosing 
90 and 50 percent of the Petrosian flux; 
\item the $^{0.1}(u-r)$ colour\footnote{We use model instead of Petrosian magnitudes to 
compute colours, since aperture photometry may include non-negligible Poisson and 
background subtraction uncertainties in the $u$ band.};
\item the stellar mass, $M_{\ast}$, computed from the absolute magnitude and colour, 
following \citet{Taylor:2011}; 
\item the probability of being elliptical ($p_E$) or spiral ($p_S$) as measured by the 
Galaxy Zoo Project (\citealt{Lintott:2011}, http://zoo1.galaxyzoo.org/).
\end{itemize}
\begin{figure*}[ht]
\centering
\includegraphics[width=18cm]{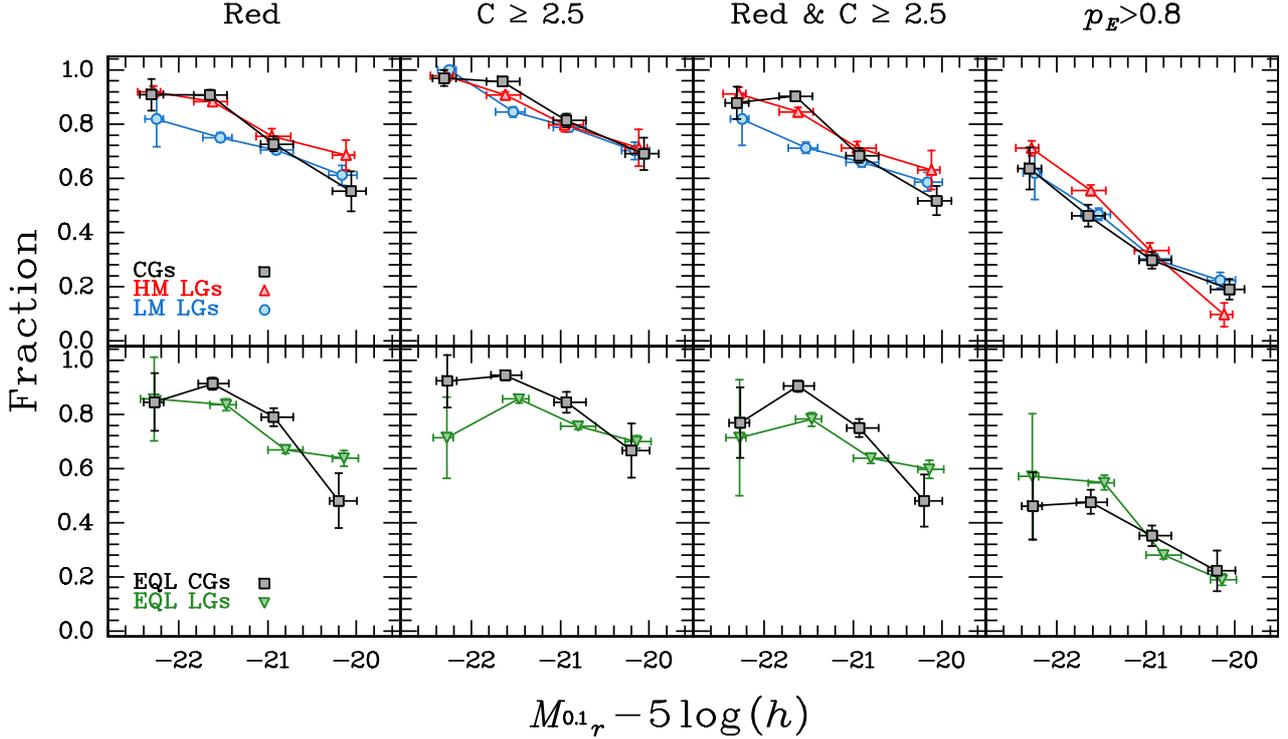}
\caption{{\em Left panels}: 
the fraction of red BCGs classified according to their $^{0.1}(u-r)$ colour; 
{\em centre left panels}: the fraction of early-type BCGs classified according to their 
concentration parameter; {\em centre right panels}: the fraction of BCGs classified 
simultaneously as red and early-type from their colour and concentration parameter; 
{\em right panels}: the fraction of BCGs classified as elliptical from their Galaxy Zoo 
morphology. All fractions are shown as a function of absolute magnitude. The {\em top 
panels} compare compact groups with loose groups of low and high mass, while {\em bottom
panels} compare compact and loose groups with similar total absolute magnitude. Vertical
error-bars are obtained by using the bootstrap resampling technique. Horizontal 
error-bars are the 25\% and 75\% quartiles of the absolute magnitude distribution within 
each bin.}
\label{fig:fraction}
\end{figure*}
Fig. \ref{fig:histo} compares the normalised distributions of BCG parameters in loose and
compact groups of galaxies. Below each panel, we show the residuals between each pair of
distributions, i.e., for each property $X$, the difference 
$\Delta F(X)=f_{CG}(X)-f_{LG}(X)$, where $f_{CG}(X)$ and $f_{LG}(X)$ are the fractions of
galaxies in the bin centred on $X$, in the compact and in the loose group sample, 
respectively. In each panel of this figure, we quote the probability for the null 
hypothesis that the parameter distributions we compare are drawn from the same 
distribution according to a Kolmogorov-Smirnov test. The best case is the comparison 
between the colours of BCGs in compact groups and in high mass loose groups, which 
reaches only a 16\% probability level. 

From the comparison of compact groups and loose groups in the low and high mass 
subsamples, we notice that there are two sets of parameters that have different 
behaviour: for some of them, the properties of the BCGs in compact groups are 
intermediate between low- and high-mass loose groups, as is the case of absolute 
magnitude, stellar mass, Petrosian half-light radii, colour and the morphology 
parameters; on the other hand, according to the two remaining parameters, surface 
brightness and concentration, the brightest galaxies of compact groups are different from
their loose-group counterparts in some more fundamental way. As an example of the first 
set, if we consider luminosity, BCGs in compact groups are typically brighter than BCGs 
in low-mass loose groups and fainter than BCGs in high-mass loose groups. That is not the
case of, for instance, their concentration: galaxies in compact groups are more 
concentrated than galaxies in loose groups, irrespective of group mass.

Comparison of groups of similar luminosity distributions ({\em bottom row} in Fig. 
\ref{fig:histo}) reveals differences between BCGs in loose and compact groups in the 
same sense as the differences between BCGs in low-mass loose groups and compact groups 
({\em top row} in Fig. \ref{fig:histo}), albeit more pronounced. Our samples of equal 
luminosity loose and compact groups have, on average, 4.7 and 3.5 galaxies that are 
brighter than the $r=17.77$ apparent magnitude limit of the SDSS' Main Galaxy Sample, 
respectively. That is, at similar group luminosity, loose groups typically have more 
galaxies than compact groups. In addition, and as we show below 
(section \ref{sec:light}), compact groups have a larger luminosity gap between the BCG 
and the second-ranked galaxy. Thus, selecting a sample of loose groups bound to have a 
luminosity distribution similar to that of a given sample of compact groups, results in a
selection of loose groups with BCGs that are systematically fainter than their 
compact-group counterparts. Since most galaxy properties are correlated to each other 
(e.g. \citealt{Blanton:2005}), the fact that the sample of equal-luminosity compact 
groups have brighter BCGs may provide an explanation for all the differences seen in the 
{\em bottom row} of Fig. \ref{fig:histo}.
\subsection{The fraction of early-type BCGs}
\label{sec:fraction}
In clusters of galaxies, BCGs are typically early-type galaxies. In our samples (see Fig.
\ref{fig:histo}), the fraction of BCGs that have a low probability of being elliptical, 
or the fraction of them that are blue, is not negligible. This subsection analyses in 
more detail the BCGs in our group samples by classifying them according to their colour,
concentration and Galaxy Zoo morphology. 

In Fig. \ref{fig:fraction} we show the fraction of BCGs as a function of galaxy absolute 
magnitude that can be classified as early-type according to four different criteria:

\begin{enumerate}
\item {\em Colour}: we consider that a galaxy belongs to the red sequence if its 
$^{0.1}(u-r)$ colour is redder than the luminosity dependent threshold of \citet{ZM11}.
\item {\em Concentration parameter}:
\citet{Strateva:2001} found that the concentration parameter can be used to differentiate
between early and late-types. We use the $r$-band concentration index, and consider as 
early-type galaxies those that have $C \ge 2.5$.
\item {\em Colour and concentration}:  We select galaxies that are redder than the 
luminosity dependent threshold of \citet{ZM11} and have $C \ge 2.5$.
\item {\em Galaxy Zoo morphology}: in addition, we use the morphological classiﬁcations 
taken from the Galaxy Zoo Project (http://zoo1.galaxyzoo.org/, \citealt{Lintott:2011})
to classify galaxies as elliptical if $p_E>0.8$. We use the $p_E$ values corrected after 
the debiasing procedure of \citet{Lintott:2011}. 
\end{enumerate}

The fraction of BCGs that we classify as early-type according to the criteria i$-$iii is
similar between compact groups and high mass loose groups and smaller for the low mass 
sample. Colour is the parameter that distinguishes best between compact groups and 
low-mass loose groups; classifying galaxies as early-type according to colour and 
concentration does not give any further information. Thus, according to these results, 
the fraction of early-type BCGs in compact groups resembles that of the massive loose 
groups. Comparing groups with similar luminosities shows that compact groups have a 
larger fraction of early-type galaxies. Within the fraction of BCGs that have a high 
probability of being elliptical according to the Galaxy Zoo Project (criterion iv), we 
notice differences: the trend for compact groups is similar to that of low-mass loose 
groups, and, within errors, the trends for equal-luminosity samples are comparable to 
each other.
\subsection{The fraction of light in the BCG}
\label{sec:light}

\begin{figure}[t]
\centering
\includegraphics[width=9cm]{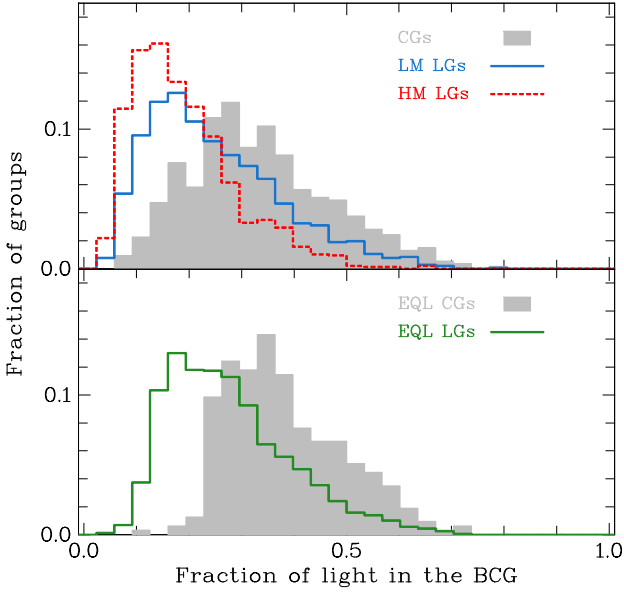}
\caption{
The distributions of the fraction of group total luminosity contained in the BCG in our 
samples.
}
\label{fig:flum}
\end{figure}

Fig. \ref{fig:flum} shows the distributions of the fraction of group light contained
in the brightest galaxy. It is clear from this figure that the brightest group galaxies 
in compact groups differ markedly from their loose-group counterparts. While they may 
not be as bright as some BCGs in massive loose groups, in terms of luminosity BCGs in 
compact groups have a more dominant position within their system.
Another conclusion from Fig. \ref{fig:flum} is that, in terms of the fraction of light 
they contain, BCGs in loose groups become less important the more massive the systems 
that are considered. This agrees with the results by \citet{Loh:2006} in clusters of 
galaxies.\\
\subsection{Statistical test for BCG luminosities}
\label{sec:light}

The statistical test developed by \citet{Tremaine:1977} enables us to distinguish between
two possible scenarios for the BCGs: either they are just extreme examples of a 
luminosity function or, on the contrary, they are objects that may have had a different 
evolution compared to their companion galaxies. The test is independent of the assumed 
luminosity and of its variation from  cluster to cluster. The key insight of the test is
that it makes use of the magnitude difference between the first two ranked galaxies, 
$\Delta M_{12}$. For a luminosity function with an exponential cut-off at the bright end,
$\phi(M)\sim \exp(\alpha M)$, given that BCGs have a small magnitude spread, the 
difference $\Delta M_{12}$ cannot be excessively large if both galaxies are drawn from 
the same luminosity function.

\citet{Tremaine:1977} defined two parameters:
\begin{equation}
 T_1=\frac{s(M_1)}{\langle \Delta M_{12} \rangle}
\end{equation}
and
\begin{equation}
 T_2=\frac{s(\Delta M_{12})}{0.667^{1/2}\langle \Delta M_{12} \rangle}
\end{equation}
where $\langle \Delta M_{12} \rangle$ is the mean value of the difference 
$\Delta M_{12}$, and $s(M_1)$ and $s(\Delta M_{12})$ are the standard deviations of $M_1$
and $\Delta M_{12}$, respectively. If the first two ranked galaxies are drawn from the 
same luminosity function, then values of $T_1\gtrsim 1$ and $T_2\gtrsim 1$ should be 
expected. Values of $T_1$ and $T_2$ lower than unity imply that the first-ranked group 
galaxies are abnormally bright at the expense of the second-ranked galaxy. \\

We show in table \ref{tb:test} the mean values of $\langle \Delta M_{12}\rangle$, $T_1$ 
and $T_2$ for all our samples of groups. The $\langle \Delta M_{12} \rangle$ value is 
typically larger for compact groups, irrespective of whether the comparison is made with
loose groups of different masses or of similar luminosities. BCGs in compact groups are 
more luminous relative to their companions. Loose groups in all our samples have values 
$T_1$ and $T_2$ similar to or greater than unity, thus consistent with being objects at 
the bright extreme of the luminosity function of galaxies in groups. This is in agreement
with \citet{Geller:1983} and \citet{Lin:2010} but not with the results by 
\citet{Paranjape12}. On the other hand, $T_1$ and $T_2$ values are lower for compact 
groups, below unity but close to it, in between the values obtained by \citet{Diaz12}
using their own sample of compact groups and  the values computed by the same 
authors using the samples by \citet{Hickson:1992,Allam:2000,Focardi:2002}.
\begin{table*}[ht]
\center
\begin{tabular}{l|*{3}{c}|*{2}{c}}
                     & Compact Groups & \multicolumn{2}{c|}{Loose Groups} & 
\multicolumn{2}{c}{Equal Luminosity Samples} \\
                     &     &   Low Mass & High Mass  & Compact Groups & Loose Groups \\
\hline
\# groups            & 477           & 1414          & 1458          & 314           & 
1577          \\ 
$\langle \Delta M_{12}\rangle$ & $0.68\pm0.02$ & $0.53\pm0.01$ & $0.50\pm0.01$ & 
$0.76\pm0.03$ & $0.55\pm0.01$ \\ 
$T_1$                & $0.95\pm0.05$ & $1.08\pm0.03$ & $1.17\pm0.03$ & $0.74\pm0.04$ & 
$0.96\pm0.02$ \\
$T_2$                & $0.94\pm0.04$ & $1.02\pm0.02$ & $1.01\pm0.02$ & $0.88\pm0.04$ & 
$1.02\pm0.02$ \\
\end{tabular}
\caption{
The mean difference in absolute magnitude between the BCG and the 2$^{\rm nd}$ ranked 
galaxy and the $T_1$ and $T_2$ statistics \citep{Tremaine:1977} for our samples of BCGs. 
Quoted errors were computed using the bootstrap resampling technique.}
\label{tb:test}
\end{table*}
\subsection{Dry mergers \& BCGs}
The previous subsections showed some evidence that BCGs in compact groups are perhaps not
just objects in the extreme bright-end of the luminosity function of galaxies, and that 
their contribution to the parent group total luminosity is larger than the contribution 
of loose groups' BCGs to their own systems.
In this subsection, we explore whether the BCGs in compact groups have grown brighter at
the expense of some of their companions, by means of a toy model in which we take loose 
groups and brighten their BCGs by dry-merging them with another galaxy in the group. We 
will use the term {\em dry merger} to name mergers in which there is no star formation, 
regardless of whether the galaxies involved are early or late types. In our model, even 
when one or both of the galaxies undergoing a merging process may still have gas 
available, we will assume for simplicity's sake that there is no star formation. For this
purpose, we use samples of groups selected as having similar luminosity distributions,
in order to perform comparisons between systems sharing a similar physical magnitude.

In this very simple scheme, for every group in the EQL LG sample, we model the dry 
merging of its BCG with the $i-$th ranked galaxy by summing their $r-$band luminosities, 
in order to brighten the BCG at the expense of that companion. We also consider the 
possibility of stellar mass losses in the merger process by subtracting a randomly 
chosen fraction of the fainter galaxy's luminosity. This fraction is allowed to be as 
high as 30\%. Even if the stars lost during the merger process remain at the bottom 
of the potential well or form a halo around the BCG, the standard photometry of the SDSS
cannot account for this excess of light (see \citealt{Tal:2011}). Therefore, in our toy 
model we assume that the star loss during a merger also represents a loss in total group
luminosity.

\begin{figure}[h]
\centering
\includegraphics[width=9cm]{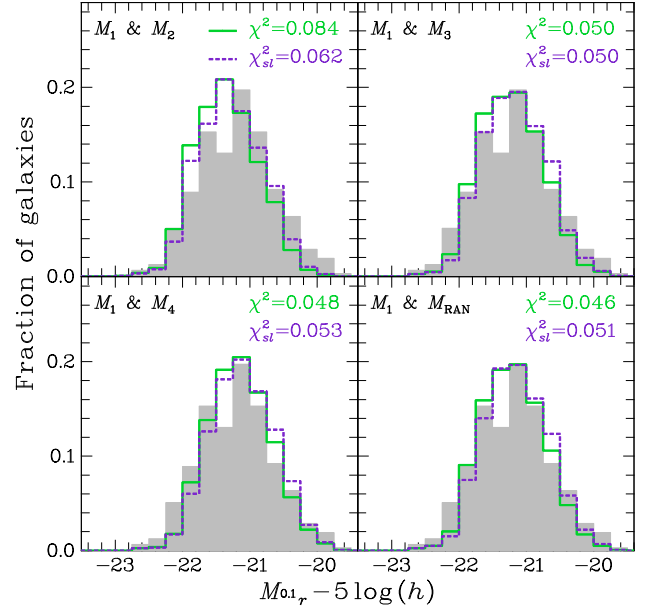}
\caption{The distributions of BCG absolute magnitudes. {\em Shaded areas:} BCGs in the 
EQL CG sample. We show in the {\em continuous green} line the distribution of absolute 
magnitudes resulting from the combined luminosity of the EQL LG BCG and: the second 
brightest galaxy in the group ({\em top left panel}); the third brightest galaxy ({\em 
top right panel}); the fourth brightest galaxy ({\em bottom left panel}); and a galaxy 
randomly chosen among the 2nd-, 3rd- and 4th-ranked galaxies. The distributions shown in the {\em dashed violet} line, are similar to the {\em green} ones, but we allow for an up
to 30\% randomly chosen fraction of stars to be lost in the merging process. We 
quote in each panel the $\chi^2$ statistics of the comparison between the models and the
{\em shaded} distribution.
}
\label{fig:merge}
\end{figure}

Fig. \ref{fig:merge} shows the absolute magnitude distribution of the galaxies resulting
from the dry merging of the BCG ($M_1$) with, alternatively, the 2nd ($M_2$), 3rd ($M_3$)
and 4th ($M_4$) ranked galaxy. We also show the result of merging the BCG with a galaxy 
randomly chosen ($M_{RAN}$) among $M_2$, $M_3$ and $M_4$. In all cases, we show the 
results with and without star loss and quote the $\chi^2$ values that result from 
comparing the model with the sample of EQL CGs. When no star loss is considered, the best
match between the absolute magnitude distribution of the model galaxies and the EQL CG 
BCGs is obtained when merging the BCG with the fourth-ranked galaxy. When we allow up to
30\% star loss in the merging, the best match occurs when the BCG merges with the 
third-ranked galaxy.

A detailed characterisation of the processes involved in the formation of the 
brightest galaxies in compact groups is far beyond the scope of our model. However, we 
can infer from the model that BCGs in compact groups may have had more mergers in their 
past history than their counterparts in loose groups, regardless of which of the former 
group members were involved in the merging process.
\section{Discussion and conclusions}
\label{sec:discussion}

We study the properties of the brightest galaxies in compact and loose groups of galaxies
to deepen our understanding of the physical mechanisms acting upon galaxy evolution in 
different environments.

We select samples of BCGs in compact groups drawn from \citet{McConnachie:2009}, and in 
loose groups taken from \citet{ZM11}. A number of physical properties of the BCGs are 
compared in compact groups and in subsamples of loose groups defined by their mass and 
total luminosity, namely: absolute magnitude, colour, size, surface brightness, stellar 
mass, concentration as well as morphological information from the Galaxy Zoo. We also 
study the fraction of BCGs that are classified as red and/or early-type as a function of 
galaxy luminosity. We analyse the fraction of the group's total luminosity contained in 
the BCG and the difference in luminosity between the brightest and the second-ranked 
galaxies.

Some properties of the BCGs in compact groups are comparable to those of BCGs in average 
loose groups. However, BCGs in compact groups are systematically more concentrated and 
have a larger surface brightness than their counterparts in both high- and low-mass 
loose groups. The fractions of red and early-type BCGs in compact and high mass loose 
groups are consistent with each other. Comparing BCGs in subsamples of compact and loose
groups selected to have similar luminosities, we find that BCGs in compact groups are, on
average, brighter, more massive, larger, redder and more frequently classified as 
elliptical. 

Compared to BCGs in loose groups, BCGs in compact groups are found to contain a larger 
fraction of the system's total luminosity and differ more in absolute magnitude from the 
second-ranked galaxy. Using a simple model, in which we dry-merge the BCG in loose groups
with another, randomly chosen, galaxy in the group, and allowing for some star lossin the process, we show that the absolute magnitude distributions of BCGs in compact and
loose groups of similar luminosities can be made more alike.

We have shown in a previous work (Paper I), that the overall galaxy population in compact
groups has undergone a major transformation compared to loose-group galaxies. In this 
work we find that their BCGs also differ. Some mechanisms responsible for transforming 
late-type galaxies into early types, such as mergers, may be more effective within 
compact groups due to their high densities and small velocity dispersion, thus leading 
their BCGs along somewhat different evolutionary paths.

From our analyses of the fraction of group light in the BCG and the difference in 
absolute magnitude relative to the second ranked galaxy, it is clear that BCGs in compact
groups are more luminous compared to their companions or to their parent group than BCGs 
in loose groups.

We find values of the Tremaine \& Richstone statistics for compact groups that are 
slightly below unity, which may indicate that their BCGs are not completely consistent 
with being part of the bright end of the luminosity function for galaxies in groups. 
Our measurements of $T_1$ and $T_2$ are in between values found in previous works, 
below unity but close to it, larger than the values obtained by \citet{Diaz12}, and 
smaller than, although closer to, the results the same authors obtained from the 
samples by \citet{Hickson:1992}, \citet{Allam:2000}, and \citet{Focardi:2002}. 

On the other hand, BCGs in our loose-group samples are clearly drawn from the bright end
of the luminosity function, in agreement with \citet{Geller:1983}. These facts may be an 
indication of the different evolution of central galaxies in compact groups.

In a hierarchical scenario for galaxy formation and evolution, mergers play a key role in
the building up of galaxies. Galaxies get bigger by accretion and by merging with other 
galaxies. Given their isolation, compact groups are not likely to have had much available
material in their surroundings to accrete. In particular, to explain their prominent 
luminosity relative to the system they inhabit, BCGs in compact groups may have had a 
merger history that contributed very efficiently to their growth in luminosity/mass 
compared to their companion galaxies. Mergers that built up the central galaxies in 
compact groups may have been more efficient per unit of group luminosity/mass than 
mergers that originated the central galaxies in loose groups.
\begin{acknowledgements}
This work was supported with grants from CONICET (PIP 11220080102603 and 11220100100336);
Ministerio de Ciencia y Tecnolog\'ia de la Provincia de C\'ordoba, Argentina 
(PID 2008/14797627); and Secretar\'ia de Ciencia y Tecnolog\'ia Universidad Nacional de 
C\'ordoba, Argentina. Funding for the Sloan Digital Sky Survey (SDSS) was provided by 
the Alfred P. Sloan 
Foundation, the Participating Institutions, the National Aeronautics and Space 
Administration, the National Science Foundation, the U.S. Department of Energy, 
the Japanese Monbukagakusho, and the Max Planck Society. The SDSS Web site is 
http://www.sdss.org/.
The SDSS is managed by the Astrophysical Research Consortium (ARC) for the 
Participating Institutions. The Participating Institutions are The University 
of Chicago, Fermilab, the Institute for Advanced Study, the Japan Participation 
Group, The Johns Hopkins University, the Korean Scientist Group, Los Alamos 
National Laboratory, the Max Planck Institut f\"ur Astronomie (MPIA), the 
Max Planck Institut f\"ur Astrophysik (MPA), New Mexico State University, 
University of Pittsburgh, University of Portsmouth, Princeton University, 
the United States Naval Observatory, and the University of Washington.
\end{acknowledgements}
\bibliographystyle{aa} 
\bibliography{ms} 

\begin{thebibliography}{43}
\expandafter\ifx\csname natexlab\endcsname\relax\def\natexlab#1{#1}\fi

\bibitem[{{Abazajian} {et~al.}(2009){Abazajian}, {Adelman-McCarthy},
  {Ag{\"u}eros}, {Allam}, {Allende Prieto}, {An}, {Anderson}, {Anderson},
  {Annis}, {Bahcall}, \& et~al.}]{dr7}
{Abazajian}, K.~N., {Adelman-McCarthy}, J.~K., {Ag{\"u}eros}, M.~A., {et~al.}
  2009, \apjs, 182, 543

\bibitem[{{Adelman-McCarthy} {et~al.}(2008){Adelman-McCarthy}, {Ag{\"u}eros},
  {Allam}, {Allende Prieto}, {Anderson}, {Anderson}, {Annis}, {Bahcall},
  {Bailer-Jones}, {Baldry}, {Barentine}, \& et~al.}]{dr6}
{Adelman-McCarthy}, J.~K., {Ag{\"u}eros}, M.~A., {Allam}, S.~S., {et~al.} 2008,
  \apjs, 175, 297

\bibitem[{{Allam} \& {Tucker}(2000)}]{Allam:2000}
{Allam}, S.~S. \& {Tucker}, D.~L. 2000, Astronomische Nachrichten, 321, 101

\bibitem[{{Beers} {et~al.}(1990){Beers}, {Flynn}, \& {Gebhardt}}]{Beers:1990}
{Beers}, T.~C., {Flynn}, K., \& {Gebhardt}, K. 1990, \aj, 100, 32

\bibitem[{{Blanton} {et~al.}(2003){Blanton}, {Brinkmann}, {Csabai}, {Doi},
  {Eisenstein}, {Fukugita}, {Gunn}, {Hogg}, \& {Schlegel}}]{Blanton:2003}
{Blanton}, M.~R., {Brinkmann}, J., {Csabai}, I., {et~al.} 2003, \aj, 125, 2348

\bibitem[{{Blanton} {et~al.}(2005){Blanton}, {Eisenstein}, {Hogg}, {Schlegel},
  \& {Brinkmann}}]{Blanton:2005}
{Blanton}, M.~R., {Eisenstein}, D., {Hogg}, D.~W., {Schlegel}, D.~J., \&
  {Brinkmann}, J. 2005, \apj, 629, 143

\bibitem[{{Coenda} {et~al.}(2012){Coenda}, {Muriel}, \&
  {Mart{\'{\i}}nez}}]{PaperI}
{Coenda}, V., {Muriel}, H., \& {Mart{\'{\i}}nez}, H.~J. 2012, \aap, 543, A119

\bibitem[{{Collins} {et~al.}(2009){Collins}, {Stott}, {Hilton}, {Kay},
  {Stanford}, {Davidson}, {Hosmer}, {Hoyle}, {Liddle}, {Lloyd-Davies}, {Mann},
  {Mehrtens}, {Miller}, {Nichol}, {Romer}, {Sahl{\'e}n}, {Viana}, \&
  {West}}]{Collins09}
{Collins}, C.~A., {Stott}, J.~P., {Hilton}, M., {et~al.} 2009, \nat, 458, 603

\bibitem[{{Conroy} {et~al.}(2007){Conroy}, {Wechsler}, \&
  {Kravtsov}}]{Conroy07}
{Conroy}, C., {Wechsler}, R.~H., \& {Kravtsov}, A.~V. 2007, \apj, 668, 826

\bibitem[{{De Lucia} \& {Blaizot}(2007)}]{DeLucia07}
{De Lucia}, G. \& {Blaizot}, J. 2007, \mnras, 375, 2

\bibitem[{{D{\'{\i}}az-Gim{\'e}nez} {et~al.}(2012){D{\'{\i}}az-Gim{\'e}nez},
  {Mamon}, {Pacheco}, {Mendes de Oliveira}, \& {Alonso}}]{Diaz12}
{D{\'{\i}}az-Gim{\'e}nez}, E., {Mamon}, G.~A., {Pacheco}, M., {Mendes de
  Oliveira}, C., \& {Alonso}, M.~V. 2012, \mnras, 426, 296

\bibitem[{{Edwards} \& {Patton}(2012)}]{Edwards12}
{Edwards}, L.~O.~V. \& {Patton}, D.~R. 2012, \mnras, 425, 287

\bibitem[{{Focardi} \& {Kelm}(2002)}]{Focardi:2002}
{Focardi}, P. \& {Kelm}, B. 2002, \aap, 391, 35

\bibitem[{{Geller} \& {Postman}(1983)}]{Geller:1983}
{Geller}, M.~J. \& {Postman}, M. 1983, \apj, 274, 31

\bibitem[{{Hickson}(1982)}]{Hickson:1982}
{Hickson}, P. 1982, \apj, 255, 382

\bibitem[{{Hickson} {et~al.}(1992{\natexlab{a}}){Hickson}, {Mendes de
  Oliveira}, {Huchra}, \& {Palumbo}}]{Hickson92}
{Hickson}, P., {Mendes de Oliveira}, C., {Huchra}, J.~P., \& {Palumbo}, G.~G.
  1992{\natexlab{a}}, \apj, 399, 353

\bibitem[{{Hickson} {et~al.}(1992{\natexlab{b}}){Hickson}, {Mendes de
  Oliveira}, {Huchra}, \& {Palumbo}}]{Hickson:1992}
{Hickson}, P., {Mendes de Oliveira}, C., {Huchra}, J.~P., \& {Palumbo}, G.~G.
  1992{\natexlab{b}}, \apj, 399, 353

\bibitem[{{Huchra} \& {Geller}(1982)}]{H&G:1982}
{Huchra}, J.~P. \& {Geller}, M.~J. 1982, \apj, 257, 423

\bibitem[{{Lidman} {et~al.}(2012){Lidman}, {Suherli}, {Muzzin}, {Wilson},
  {Demarco}, {Brough}, {Rettura}, {Cox}, {DeGroot}, {Yee}, {Gilbank},
  {Hoekstra}, {Balogh}, {Ellingson}, {Hicks}, {Nantais}, {Noble}, {Lacy},
  {Surace}, \& {Webb}}]{Lidman12}
{Lidman}, C., {Suherli}, J., {Muzzin}, A., {et~al.} 2012, \mnras, 427, 550

\bibitem[{{Limber} \& {Mathews}(1960)}]{Limber:1960}
{Limber}, D.~N. \& {Mathews}, W.~G. 1960, \apj, 132, 286

\bibitem[{{Lin} {et~al.}(2010){Lin}, {Ostriker}, \& {Miller}}]{Lin:2010}
{Lin}, Y.-T., {Ostriker}, J.~P., \& {Miller}, C.~J. 2010, \apj, 715, 1486

\bibitem[{{Lintott} {et~al.}(2011){Lintott}, {Schawinski}, {Bamford}, {Slosar},
  {Land}, {Thomas}, {Edmondson}, {Masters}, {Nichol}, {Raddick}, {Szalay},
  {Andreescu}, {Murray}, \& {Vandenberg}}]{Lintott:2011}
{Lintott}, C., {Schawinski}, K., {Bamford}, S., {et~al.} 2011, \mnras, 410, 166

\bibitem[{{Liu} {et~al.}(2009){Liu}, {Mao}, {Deng}, {Xia}, \& {Wen}}]{Liu09}
{Liu}, F.~S., {Mao}, S., {Deng}, Z.~G., {Xia}, X.~Y., \& {Wen}, Z.~L. 2009,
  \mnras, 396, 2003

\bibitem[{{Loh} \& {Strauss}(2006)}]{Loh:2006}
{Loh}, Y.-S. \& {Strauss}, M.~A. 2006, \mnras, 366, 373

\bibitem[{{Mart{\'{\i}}nez} \& {Muriel}(2006)}]{MM2:2006}
{Mart{\'{\i}}nez}, H.~J. \& {Muriel}, H. 2006, \mnras, 370, 1003

\bibitem[{{Mart{\'{\i}}nez} \& {Zandivarez}(2012)}]{Martinez:2012}
{Mart{\'{\i}}nez}, H.~J. \& {Zandivarez}, A. 2012, \mnras, 419, L24

\bibitem[{{Martizzi} {et~al.}(2012){Martizzi}, {Teyssier}, \&
  {Moore}}]{Martizzi12}
{Martizzi}, D., {Teyssier}, R., \& {Moore}, B. 2012, \mnras, 420, 2859

\bibitem[{{McConnachie} {et~al.}(2009){McConnachie}, {Patton}, {Ellison}, \&
  {Simard}}]{McConnachie:2009}
{McConnachie}, A.~W., {Patton}, D.~R., {Ellison}, S.~L., \& {Simard}, L. 2009,
  \mnras, 395, 255

\bibitem[{{Moore} {et~al.}(1993){Moore}, {Frenk}, \& {White}}]{Moore:1993}
{Moore}, B., {Frenk}, C.~S., \& {White}, S.~D.~M. 1993, \mnras, 261, 827

\bibitem[{{Paranjape} \& {Sheth}(2012)}]{Paranjape12}
{Paranjape}, A. \& {Sheth}, R.~K. 2012, \mnras, 423, 1845

\bibitem[{{Petrosian}(1976)}]{petro76}
{Petrosian}, V. 1976, \apjl, 209, L1

\bibitem[{{Postman} \& {Lauer}(1995)}]{postman:1995}
{Postman}, M. \& {Lauer}, T.~R. 1995, \apj, 440, 28

\bibitem[{{Schlegel} {et~al.}(1998){Schlegel}, {Finkbeiner}, \&
  {Davis}}]{sch98}
{Schlegel}, D.~J., {Finkbeiner}, D.~P., \& {Davis}, M. 1998, \apj, 500, 525

\bibitem[{{Smith} {et~al.}(2010){Smith}, {Khosroshahi}, {Dariush}, {Sanderson},
  {Ponman}, {Stott}, {Haines}, {Egami}, \& {Stark}}]{Smith10}
{Smith}, G.~P., {Khosroshahi}, H.~G., {Dariush}, A., {et~al.} 2010, \mnras,
  409, 169

\bibitem[{{Stott} {et~al.}(2010){Stott}, {Collins}, {Sahl{\'e}n}, {Hilton},
  {Lloyd-Davies}, {Capozzi}, {Hosmer}, {Liddle}, {Mehrtens}, {Miller}, {Romer},
  {Stanford}, {Viana}, {Davidson}, {Hoyle}, {Kay}, \& {Nichol}}]{Stott10}
{Stott}, J.~P., {Collins}, C.~A., {Sahl{\'e}n}, M., {et~al.} 2010, \apj, 718,
  23

\bibitem[{{Strateva} {et~al.}(2001){Strateva}, {Ivezi{\'c}}, {Knapp},
  {Narayanan}, {Strauss}, {Gunn}, {Lupton}, {Schlegel}, {Bahcall}, \&
  et~al.}]{Strateva:2001}
{Strateva}, I., {Ivezi{\'c}}, {\v Z}., {Knapp}, G.~R., {et~al.} 2001, \aj, 122,
  1861

\bibitem[{{Strauss} {et~al.}(2002){Strauss}, {Weinberg}, {Lupton}, {Narayanan},
  {Annis}, {Bernardi}, {Blanton}, {Burles}, {Connolly}, {Dalcanton}, {Doi},
  {Eisenstein}, \& et~al.}]{Strauss:2002}
{Strauss}, M.~A., {Weinberg}, D.~H., {Lupton}, R.~H., {et~al.} 2002, \aj, 124,
  1810

\bibitem[{{Tal} \& {van Dokkum}(2011)}]{Tal:2011}
{Tal}, T. \& {van Dokkum}, P.~G. 2011, \apj, 731, 89

\bibitem[{{Taylor} {et~al.}(2011){Taylor}, {Hopkins}, {Baldry}, {Brown},
  {Driver}, {Kelvin}, {Hill}, {Robotham}, {Bland-Hawthorn}, {Jones}, {Sharp},
  {Thomas}, {Liske}, {Loveday}, {Norberg}, {Peacock}, {Bamford}, {Brough},
  {Colless}, {Cameron}, {Conselice}, {Croom}, {Frenk}, {Gunawardhana},
  {Kuijken}, {Nichol}, {Parkinson}, {Phillipps}, {Pimbblet}, {Popescu},
  {Prescott}, {Sutherland}, {Tuffs}, {van Kampen}, \&
  {Wijesinghe}}]{Taylor:2011}
{Taylor}, E.~N., {Hopkins}, A.~M., {Baldry}, I.~K., {et~al.} 2011, \mnras, 418,
  1587

\bibitem[{{Tonini} {et~al.}(2012){Tonini}, {Bernyk}, {Croton}, {Maraston}, \&
  {Thomas}}]{Tonini12}
{Tonini}, C., {Bernyk}, M., {Croton}, D., {Maraston}, C., \& {Thomas}, D. 2012,
  \apj, 759, 43

\bibitem[{{Tremaine} \& {Richstone}(1977)}]{Tremaine:1977}
{Tremaine}, S.~D. \& {Richstone}, D.~O. 1977, \apj, 212, 311

\bibitem[{{Whiley} {et~al.}(2008){Whiley}, {Arag{\'o}n-Salamanca}, {De Lucia},
  {von der Linden}, {Bamford}, {Best}, {Bremer}, {Jablonka}, {Johnson},
  {Milvang-Jensen}, {Noll}, {Poggianti}, {Rudnick}, {Saglia}, {White}, \&
  {Zaritsky}}]{Whiley08}
{Whiley}, I.~M., {Arag{\'o}n-Salamanca}, A., {De Lucia}, G., {et~al.} 2008,
  \mnras, 387, 1253

\bibitem[{{Zandivarez} \& {Mart{\'{\i}}nez}(2011)}]{ZM11}
{Zandivarez}, A. \& {Mart{\'{\i}}nez}, H.~J. 2011, \mnras, 415, 2553

\end{thebibliography}
\end{document}